# Chiral phase-imaging meta-sensors

Ahmet M. Erturan†, Jianing Liu†, Maliheh A. Roueini†, Nicolas Malamug,

Lei Tian, and Roberto Paiella

Department of Electrical and Computer Engineering and Photonics Center,

Boston University, 8 Saint Mary's Street, Boston, MA 02215

*† These authors contributed equally*

ABSTRACT: Light waves possess multiple degrees of freedom besides intensity, including phase and polarization, that often contain important information but require complex and bulky systems for their measurement. Here we report a pair of compact multifunctional photodetectors that can selectively measure the local phase gradient of, respectively, the right and left circular-polarization component of any incident wave. These devices employ a chiral pair of integrated plasmonic metasurfaces to introduce a sharp dependence of responsivity on local direction of propagation of the desired polarization component. An order-of-magnitude polarization selectivity with respect to phase gradient is demonstrated with both devices. Using the measured device characteristics, we also describe computationally a pixel array that allows for the simultaneous separate mapping of the right and left circularly-polarized incident wavefronts in a particularly simple imaging setup. These unique capabilities may be exploited to enable new functionalities for applications in chemical sensing, biomedical microscopy, and machine vision.

# 1. Introduction

Traditional photodetectors are designed to capture all the light impinging on their illumination window and convert it into an electrical signal proportional to its optical intensity. Such devices are used routinely in countless light-sensing applications in science, technology, and everyday life. At the same time, however, their basic operation principle does not allow for the direct detection of all the other properties of the incident light (phase, polarization, wavelength, angular momentum), which can also contain important information of interest. Instead, the measurement of these properties generally requires the use of complex and bulky optical setups (e.g., based on interferometry, polarimetry, spectral or spatial filtering), which limits their widespread applicability. In recent years, the development of more functional and miniaturized photodetectors capable of measuring multiple degrees of freedom simultaneously has thus emerged as a major research goal [1]. Broadly speaking, these devices rely on recent advances in nanophotonics and nanomaterials to encode the incident-light properties of interest on their output readout signals. The desired information is then retrieved computationally by decoding these signals with various numerical techniques.

An important example of such multifunctional light sensors is photodetectors featuring a sharp dependence of responsivity on illumination angles [2-6]. When combined in pixel arrays, these devices can be used to map the incident wavefronts, and thus the spatial distribution of the optical phase, for applications such as surface profiling and phase contrast imaging of transparent biological cells [7]. Polarization-sensitive photodetectors have also been the subject of extensive recent work, particularly focused on the use of chiral nanomaterials [8, 9] and nanostructures [10-13] for the selective intensity detection of right or left circular polarization (RCP or LCP) without external polarizers and wave plates. In turn, the use of circularly polarized light provides capabilities of interest for the spectroscopic detection of biomolecules [14], imaging of chiral materials [15], quantum science and information processing [16], and remote sensing [17,18].

In this general context, the present work introduces an entirely new device functionality – the ability to selectively measure the local phase gradient of only one circular polarization component of the incident light. This behavior is enabled by a chiral plasmonic metasurface supported by a metal film on the illumination window of a planar photodetector [Fig. 1(a)]. Depending on the metasurface design, the RCP or LCP component of the incident light is coupled into surface plasmon polaritons (SPPs) guided by the metal film. These excited SPPs are then scattered into the photodetector active layer by a nearby set of slits perforated through the metal film. The incident-light-to-SPP coupling is determined by a phase matching condition that makes the metasurface transmission (and thus the device responsivity) strongly dependent on illumination angle near normal incidence [see red trace in the simulation results of Fig. 1(b)]. As a result, a small deflection in the direction of propagation (i.e., transverse phase gradient) for the chosen polarization component produces a large change in the device photocurrent signal. This change can be measured with a differential detection scheme, where the signals of two adjacent oppositely-oriented devices in a pixel array are subtracted from each other. At the same time, the angular response for the other circular polarization component [blue trace in Fig. 1(b)] is weaker and nearly isotropic (aside from small fluctuations), so that its contribution to the same differential measurement cancels out.

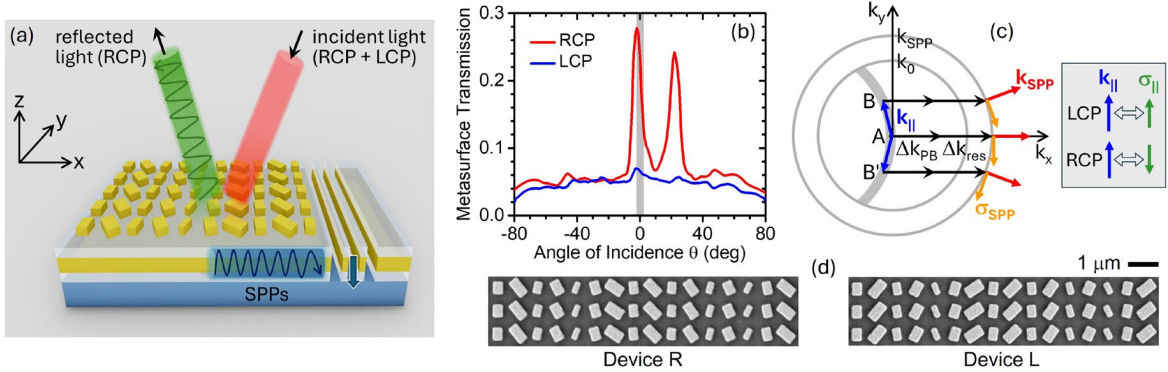

**Figure 1.** Chiral metasurface photodetectors. (a) Schematic device structure. At the target detection angle, one circular polarization component of the incident light is reflected, while the other is coupled to SPPs. (b) Calculated transmission through the device metasurface for RCP and LCP light at $\lambda_0 = 1550$ nm versus angle of incidence $\theta$ on the x–z plane. The shaded region shows the range of possible angles of incidence on the sensor array for a representative microscope configuration with 0.8 objective numerical aperture and 20× magnification. (c) Reciprocal-space diagram illustrating the plasmon excitation process in these devices, for light incident along 3 representative directions (labeled A, B, and B'). The red and orange arrows represent the wavevector $\mathbf{k}_{SPP}$ and spin angular momentum $\boldsymbol{\sigma}_{SPP}$ of the excited SPPs. The combined phase-matching action of the metasurface resonance and PB phase is indicated by the horizontal black arrows. The panel on the right-hand side shows the directional relation between in-plane wavevector $\mathbf{k}_\parallel$ (blue arrow) and spin $\boldsymbol{\sigma}_\parallel$ (green arrow) for LCP and RCP light. The SPP excitation efficiency is enhanced (suppressed) when $\boldsymbol{\sigma}_\parallel$ is parallel (antiparallel) to $\boldsymbol{\sigma}_{SPP}$. (d) Top-view SEM images of the metasurfaces in the two devices reported in this work, illustrating their chiral relationship.

To demonstrate these ideas, here we report the design, fabrication and characterization of a pair of metasurface photodetectors of opposite chirality allowing for phase contrast sensing of circularly polarized light of opposite handedness. The expected dependence of responsivity on illumination angle and polarization is clearly observed in both devices via photocurrent measurements. The experimental results also provide a vivid illustration of the plasmonic spin-Hall effect [19], whereby the measured angle- and polarization-resolved responsivity further depends on the spin-angular-momentum matching between the incident circularly polarized light and the excited SPPs. Finally, we present computational imaging simulations based on the measured angular response maps to show how these devices can be combined in a pixel array for the simultaneous and independent mapping of the incident RCP and LCP wavefronts, in a particularly compact single-shot imaging setup. It should be noted that, while metasurface flat optics has been used in recent years for phase imaging [6, 20-24] and for circular polarization filtering [11, 25-27], the combination of both functionalities in a single device architecture is unique to the present work.

These novel phase-imaging meta-sensors (PIMSs) could find immediate use in multiple imaging applications involving chiral matter, e.g., for drug development [28] (given the prevalence of chiral chemicals in pharmaceutical substances), fundamental studies of live cells [29], and biomedical diagnostics [30, 31]. Importantly, the same tasks currently require complex and bulky combinations of polarization and spatial-filtering or interference optics, often in conjunction with slow sequential measurement protocols. The devices reported in this work could therefore enable new sensing capabilities in situations where miniaturization and speed are important (e.g., for endoscopy and *in vivo* microscopy). The combination of phase and polarization imaging in a simple and compact setup is also quite promising for navigation and remote sensing under conditions of limited intensity contrast [17, 18]. In this respect, it should be noted that one of the most advanced vision systems in nature (the compound eye of the mantis shrimp) utilizes the ability to discriminate between RCP and LCP light for object identification and contrast enhancement [32]. More broadly, these chiral PIMSs provide a key missing ingredient for the development of a sensor array that could directly measure, in a single shot, the spatial distribution of the incident optical intensity, phase, and polarization. Such a system would add a major new functionality (i.e., the ability to map polarization-resolved phase contrast) to existing image-sensor technologies, including recently developed metasurface full-Stokes polarization cameras [33-35].

As a result, it could dramatically improve our ability to sense and process visual information in challenging environments for countless applications of machine vision.

## 2. Results

## 2.1 Metasurface design

The metasurfaces developed in this work consist of Au rectangular nanoparticles (NPs) of different dimensions and orientations, arranged in a square lattice on a $SiO_2/Au/SiO_2$ stack deposited on the illumination window of photodetector (here a Ge photoconductor) [Fig. 1(a)]. The metal film prevents incident light from being transmitted directly into the underlying device active layer. As a result, photodetection can only take place through a plasmon-assisted process, where light incident at a desired angle $\theta_p$ and polarization (RCP or LCP) is first converted into SPPs on the Au surface. These guided waves are then efficiently scattered into the substrate by the nearby slits [36, 37], in a process analogous to extraordinary optical transmission through sub-wavelength apertures in metal films. A similar device structure, with a one-dimensional periodic diffraction grating instead of the NP metasurface array, has been developed in our prior work focused on lensless compound-eye vision [4] and linearly polarized phase contrast imaging [6]. For phase imaging, the target detection angle $\theta_p$ should be small enough so that the resulting peak in the angular response overlaps asymmetrically with normal incidence [as in Fig. 1(b)], leading to large variations in responsivity with illumination angle around $\theta = 0$.

      For the angle- and circular-polarization-selective excitation of SPPs, we rely on the combined use of the resonance phase and Pancharatnam-Berry (PB) phase of the individual meta-units, which provides a particularly effective route for chiral wavefront manipulation [38-44]. The resonance phase $\varphi_{res}$ here is associated with the excitation of localized plasmonic oscillations in

the Au NPs and thus depends on the NP lateral dimensions. Despite the dipolar nature of these resonances, $\varphi_{res}$ can be tuned over a broad range of nearly $2\pi$ through the coupling of the NP plasmonic oscillations with their mirror image in the underlying metal film [45]. The PB or geometric phase $\varphi_{PB}$ is associated with the anisotropic optical response of the rectangular NPs and thus depends on their orientation relative to the axes of the metasurface array [46]. The combined effect of both phase contributions can be evaluated from the Jones matrix that describes scattering of the incident light by each meta-unit in the RCP/LCP basis:

$$M = \frac{1}{2}\begin{bmatrix} r_\xi + r_\psi & (r_\xi - r_\psi)e^{-i2\alpha} \\ (r_\xi - r_\psi)e^{i2\alpha} & r_\xi + r_\psi \end{bmatrix}. \qquad (1)$$

Here, $r_\xi$ and $r_\psi$ are the amplitude reflection coefficients for linearly polarized light along the two axes of the rectangular NP, and $\alpha$ is the NP orientation angle relative to the sides of the square unit cell [Fig. 2(a)]. If the NPs are designed so that $|r_\xi + r_\psi| \ll |r_\xi - r_\psi|$, most of the incident RCP light experiences a reversal in the direction of field rotation accompanied by a total phase shift $\varphi_{tot} = \arg\{r_\xi - r_\psi\} + 2\alpha$. Similarly, the LCP component is scattered with the same resonance phase $\varphi_{res} = \arg\{r_\xi - r_\psi\}$ but equal and opposite PB phase $\varphi_{PB} = -2\alpha$. The selective detection of either state of circular polarization in the proposed PIMSs is enabled by this sign difference for the PB phase contribution.

Specifically, in these devices, each NP is rotated relative to its preceding unit along the x direction [in the system of coordinates of Fig. 2(a)] by a fixed angle $\Delta\alpha$. At the same time, its resonance phase (determined by its lateral dimensions $L_\xi$ and $L_\psi$) differs from that of the preceding NP by a fixed amount $\Delta\varphi_{res}$. As a result, the overall metasurface scattering phase varies linearly with NP position along the x direction with slope $\Delta k_{tot} = (\Delta\varphi_{res} \pm 2\Delta\alpha)/\Delta x$, where the plus and minus signs correspond to RCP and LCP light, respectively. Here $\Delta x$ is the array lattice constant, which we set at a reasonably small value of 600 nm, well below the PIMS design wavelength $\lambda_0$ = 1550 nm. Under these conditions, when incident light is scattered by the metasurface, its in-plane wavevector $\mathbf{k}_\parallel$ is shifted by the fixed amount $\hat{\mathbf{x}}\Delta k_{tot}$. If the resulting wavevector $\mathbf{k}_\parallel + \hat{\mathbf{x}}\Delta k_{tot}$ matches that of an SPP supported by the Au film, the same SPP can be efficiently excited by the incident light.

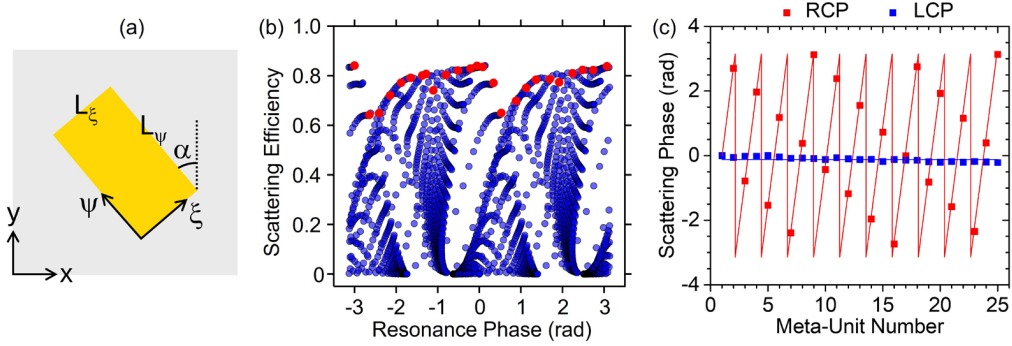

**Figure 2.** Design simulations. (a) Schematic top-view image of a meta-unit. (b) Calculated scattering efficiency $\frac{1}{4}|r_\xi - r_\psi|^2$ versus resonance phase $\varphi_{res} = \arg\{r_\xi - r_\psi\}$ for all simulated NPs. The red circles indicate the NPs used in the devices reported in this work. (c) Total scattering phase $\varphi_{tot} = \varphi_{res} + \varphi_{PB}$ of all the NPs in the metasurface of device R versus NP number for RCP (red) and LCP (blue) incident light. Device L features the same scattering phase profiles with the two states of circular polarization interchanged.

For the detection of light with $\mathbf{k}_\parallel$ along the x direction and angle of incidence $\theta_p$ [e.g., point A in the reciprocal-space diagram of Fig. 1(c)], this phase matching condition becomes $\Delta k_{tot} = k_0(n_{SPP} - \sin\theta_p)$, where $k_0 = 2\pi/\lambda_0$ and $n_{SPP}$ is the SPP effective index. To enforce this condition for RCP light, we select the NP dimensions and orientations such that $\Delta\varphi_{res}$ and $\Delta\alpha$ satisfy

$$\Delta\varphi_{res} = 2\Delta\alpha = k_0(n_{SPP} - \sin\theta_p)\Delta x/2. \qquad (2)$$

With these prescriptions, the resonance and PB phase produce equal contributions to the RCP in-plane wavevector shift $\Delta k_{tot} = (\Delta\varphi_{res} + 2\Delta\alpha)/\Delta x$, which add up to each other to bridge the $\mathbf{k}_\parallel$ mismatch $k_0(n_{SPP} - \sin\theta_p)$ between the incident light and SPPs [black arrows in Fig. 1(c)]. In contrast, under LCP illumination the resonance and PB contributions of the same metasurface become equal and opposite (due to the sign change in the latter), and therefore cancel each other to yield $\Delta k_{tot} = 0$. As a result, scattering of LCP light by the metasurface simply results in specular reflection without SPP excitation. Similarly, if $\Delta\alpha$ in Eq. (2) is replaced by $-\Delta\alpha$ (i.e., the NPs are rotated by the same angle as in the previous design but in the opposite direction), the metasurface will selectively couple LCP light incident at $\theta_p$ to SPPs while at the same time reflecting the RCP component.

Geometrically, the two metasurfaces just described form an enantiomer pair, as their NP distributions are the mirror images of one another without being superimposable. The detailed metasurface design is based on a library constructed via finite difference time domain (FDTD) simulations of individual NPs of different dimensions (see Supplementary Material, Section S1). The resulting data set is shown in Fig. 2(b), where we plot $\frac{1}{4}|r_\xi - r_\psi|^2$ versus $\varphi_{res} = \arg\{r_\xi - r_\psi\}$ at $\lambda_0$ = 1550 nm for all simulated values of $L_\xi$ and $L_\psi$ (ranging from 0 to 600 nm in steps of 10 nm). Here $\frac{1}{4}|r_\xi - r_\psi|^2$ is the cross-polarized scattering efficiency according to Eq. (1), and thus provides a measure of how efficiently the NP array can couple circularly polarized light into SPPs. The full NP arrays were constructed from this data set according to the prescriptions of Eq. (2), with $\Delta\varphi_{res}$ and $\Delta\alpha$ set to 79° and 39.5°, respectively, to produce a suitably small angle of peak detection $\theta_p$ = –2°. Each metasurface consists of 25 different NPs along the x directions, repeated periodically along y. The 25 NPs [indicated by the red circles in Fig. 2(b)] were selected to produce the required values of the resonance phase $\varphi_{res}$ while at the same time maximizing the scattering efficiency $\frac{1}{4}|r_\xi - r_\psi|^2$.

For the PIMS designed for angle-sensitive detection of RCP light (labeled device R in the following), the resulting RCP and LCP scattering phase profiles $\varphi_{tot}(x)$ are plotted in Fig. 2(c). Figure 1(b) shows FDTD simulation results for the complete device including the slits. Identical traces with the two states of circular polarization interchanged are obtained for its chiral twin (device L). As expected, the RCP metasurface transmission in Fig. 1(b) exhibits a sharp peak at the target detection angle $\theta_P = -2°$, while the LCP transmission is relatively isotropic. The shaded region in the same figure indicates the range of possible angles of incidence on the sensor array for a representative microscope configuration. Within this range, the RCP transmission varies almost linearly with angle, so that any local wavefront distortion can be measured based on the resulting change in photocurrent compared to normal-incidence illumination. The other peak observed in this trace at about 22° is associated with first-order diffraction by the NP array, whereby the incident-light wavevector is shifted by $\hat{x}(\Delta k_{tot} - 2\pi/\Delta x)$ (see Supplementary Material, Section S2). Finally, the finite background transmission and small angular fluctuations in both traces of Fig. 1(b) are ascribed to incomplete destructive interference of the light waves scattered by all the NPs in the array (due to their finite number and different scattering efficiencies), leading to partial SPP excitation even away from resonance.

## 2.2 Device fabrication and characterization

In the experimental samples, each metasurface is fabricated on a Ge substrate and surrounded by two metal contacts for biasing and current collection. The target layer thicknesses are 60/100/60 nm for the $SiO_2/Au/SiO_2$ stack supporting the NP array, and 50 nm for the Au NPs. Each slit section contains 5 linear slits perforated through the entire stack with 200-nm width and 400-nm period. The complete samples consist of a few (7) identical repetitions of the 15-μm-wide 25-NP structure described above, with each section surrounded symmetrically by two sets of slits, and with a large (300 μm) separation between the two electrodes. This arrangement (equivalent to multiple identical pixels binned together) is convenient for the angle-resolved photocurrent measurements. Figure 1(d) shows scanning electron microscopy (SEM) images of the NP arrays of the two devices. The symmetry relationship between the two metasurfaces, one being the mirror image of the other, is well evidenced in these pictures.

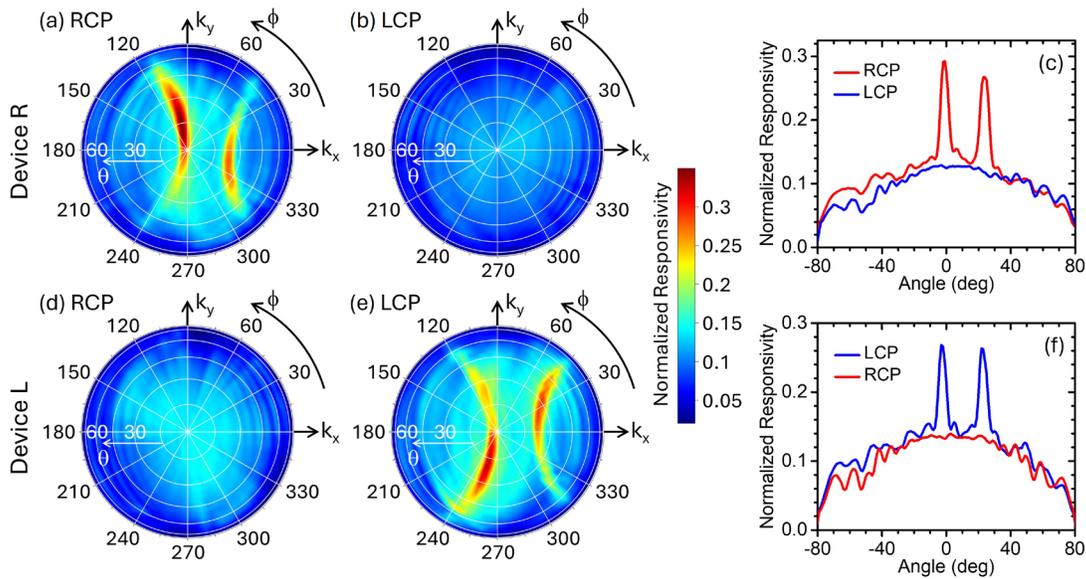

**Figure 3.** Measurement results. (a), (b) Responsivity of device R versus polar θ and azimuthal ϕ angles of incidence for RCP (a) and LCP (b) light. (c) Horizontal line cuts of the color maps of (a) and (b) (red and blue traces, respectively). (d), (e), (f) Same as (a), (b), (c) for device L. All the data presented in these plots were measured at 1550-nm wavelength and were normalized to the responsivity of a similar device without any metasurface.

The measurement results (Fig. 3) are in good agreement with expectations. The color maps of Figs. 3(a) and 3(b) show, respectively, the RCP and LCP responsivity $\mathcal{R}$ of device R measured as a function of polar θ and azimuthal ϕ angles of incidence at $\lambda_0$ = 1550 nm. The curved region of high responsivity near the origin of the RCP map is a direct consequence of the phase-matched SPP excitation process described above [see Fig. 1(c)]. The additional peak originating from first-order diffraction is also observed in the same map. Incidentally, this peak could be entirely removed from the angular response by replacing the slits on the left-hand side of the NP array with a suitably designed metasurface reflector, at the expense of a somewhat larger pixel area (responsivity data measured with such a device are presented in Supplementary Material, Section S2). In any case, for RCP-selective phase imaging, the key feature of these data is the finite versus near-zero responsivity slope $|d\mathcal{R}/d\theta|$ around normal incidence ($\theta = 0$) for RCP and LCP light, respectively [see horizontal line cuts in Fig. 3(c)]. Similar results with the two states of circular polarization interchanged were obtained with device L [Figs. 3(d)-(f)].

All the data presented in these figures are normalized to the normal-incidence responsivity of a similar device without any metasurface. The resulting peak values observed in Figs. 3(c) and 3(f) [about 29% and 27% for the RCP and LCP responsivity of devices R and L, respectively] are reasonably consistent with expectations. Specifically, from the ratio between the peak metasurface transmission computed in Fig. 1(b) (28%) and the Fresnel transmission of the uncoated Ge surface in the reference sample (62%), the expected normalized peak responsivity is 45%. The difference between the measured and calculated values can be primarily ascribed to SPP scattering by surface roughness in the experimental samples, which is also responsible for the observed decrease in peak-to-background ratio in Figs. 3(c) and 3(f) compared to Fig. 1(b).

Another distinctive feature of the experimental maps of Figs. 3(a) and 3(e) is the observed asymmetry of the curved regions of high responsivity with respect to the horizontal axis. This behavior is not accounted for by the phase-matching considerations presented above and can instead be explained in terms of spin conservation. It is now well established that SPPs possess a transverse spin angular momentum $\sigma_{SSP}$ proportional to the cross product of their real and imaginary wavevectors, which can be interpreted as a manifestation of the quantum spin Hall effect for light [47, 48]. As an illustration, in Fig. 1(c) we show the direction of $\sigma_{SSP}$ for SPPs excited in our devices by light incident with two mirror-symmetric wavevectors (at points B and B' in the $k_x$-$k_y$ plane). Circularly polarized light instead features a longitudinal spin determined by its handedness, i.e., parallel and antiparallel to the wavevector for LCP and RCP, respectively. Therefore, if LCP light is incident at point B', its in-plane spin component $\sigma_\parallel$ is nearly parallel to the spin $\sigma_{SSP}$ of the resulting excited SPP, as can be seen in Fig. 1(c). Due to spin conservation, such spin alignment maximizes the photon/SPP coupling efficiency upon scattering by each NP [49, 50]. For the same optical wave incident at B, $\sigma_\parallel$ and $\sigma_{SSP}$ are nearly antiparallel and the SPP excitation is correspondingly decreased. This argument explains the asymmetry observed in Fig. 3(e), and similar considerations in reverse can be applied to Fig. 3(a) for RCP light. The metasurface configuration reported in this work therefore also provides an interesting platform to study the role of spin conservation in photon/plasmon interactions towards novel optoelectronic device applications.

Additional measurements (shown in Supplementary Material, Section S3) reveal a steady shift in the angle of peak detection as the illumination wavelength is detuned from its design value of 1550 nm, consistent with the phase-matching condition discussed above. Figure 4(a) shows the resulting wavelength dependence of the RCP and LCP normal-incidence responsivity slopes (i.e., $|d\mathcal{R}_{RCP}/d\theta|$ and $|d\mathcal{R}_{LCP}/d\theta|$ at $\theta = 0$) of device R. For chiral phase imaging with this device, $|d\mathcal{R}_{RCP}/d\theta|$ should be as large as possible to maximize the photocurrent sensitivity to small RCP wavefront distortions; at the same time, $|d\mathcal{R}_{LCP}/d\theta|$ should be as small as possible to minimize any crosstalk from the LCP wavefronts. The plot of Fig. 4(a) therefore suggests that this sample can provide comparable phase imaging performance to the 1550-nm results presented below over a spectral range of about 70 nm (the full width at half maximum of the red trace, denoted by the double arrow). In the same spectral range, the ratio $|d\mathcal{R}_{RCP}/d\theta|/|d\mathcal{R}_{LCP}/d\theta|$ is 37× on average, which can be taken as a measure of the polarization selectivity with respect to phase gradient of this device. Similar considerations apply to device L [Fig. 4(b)], with somewhat smaller bandwidth (60 nm) but larger average selectivity (71×). It should also be noted that the operation bandwidth of these chiral PIMSs could be further expanded using more complex meta-unit geometries, e.g., similar to extensive prior work on achromatic metalenses [51].

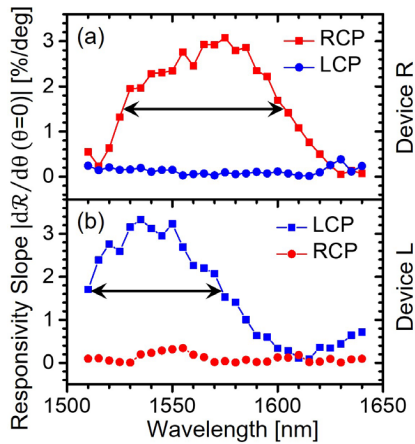

**Figure 4.** Wavelength dependence. (a) Absolute value of the normal-incidence responsivity slope under RCP (red squares) and LCP (blue circles) illumination measured as a function of wavelength with device R. (b) Same as (a) for device L. All responsivity values are normalized as in Fig. 3. The double arrows indicate the full width at half maximum of their respective traces.

## 2.3 Computational imaging results

A possible pixel configuration for the simultaneous independent measurement of the incident RCP and LCP wavefronts is shown schematically in Fig. 5. Here the sensor array is partitioned into blocks of four adjacent pixels coated with the metasurfaces of devices R, L, and their replicas rotated by 180° (labeled $\bar{R}$ and $\bar{L}$ in the following). The experimental RCP and LCP responsivity maps of all four devices in each superpixel are also shown in the figure (same as in Fig. 3 but zoomed in on a narrower region of the $k_x$-$k_y$ plane). The black circles in these plots indicate the pupil-function cutoff spatial frequency of the imaging system (corresponding to a maximum angle $\theta_c = 2.3°$), assuming 20× magnification and 0.8 objective numerical aperture. Within these circles, one responsivity map of each device exhibits a nearly linear dependence on $k_x$, whereas the other is essentially constant with **k**. As a result, any deflection in the local direction of propagation of the RCP component away from normal incidence produces equal and opposite changes in $I_R$ and $I_{\bar{R}}$ (the photocurrent signals measured by the R and $\bar{R}$ PIMSs in each superpixel), and no significant change in $I_L$ and $I_{\bar{L}}$ (and vice versa for any LCP wavefront distortion).

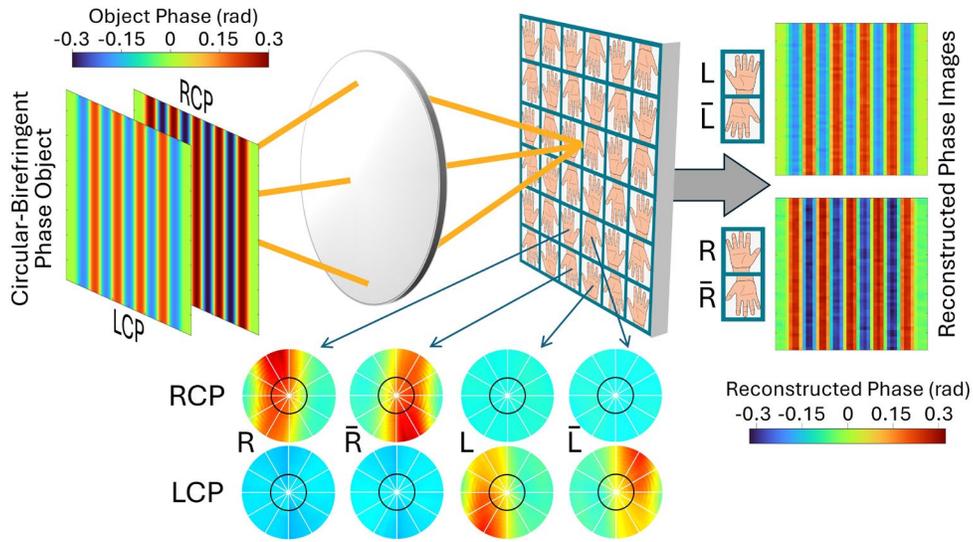

**Figure 5.** Chiral phase imaging system. The sensor array is partitioned into blocks of four adjacent pixels coated with the metasurfaces of devices R, L, and their replicas rotated by 180° (labeled $\bar{R}$ and $\bar{L}$). In this example, the incident light consists of a superposition of RCP and LCP components with equal magnitudes and different phase profiles. The RCP (LCP) phase profile can be reconstructed from the readout signals of pixels R and $\bar{R}$ (L and $\bar{L}$). The experimental RCP and LCP angular response maps of all four pixels in each block are also shown for $\theta \leq 6°$. The black circle in each color map indicates the pupil-function cutoff frequency of the imaging optics, corresponding to a maximum angle of incidence $\theta_c = 2.3°$.

The imaging system of Fig. 5 can therefore record, in a single shot, two independent differential-phase-contrast images of the RCP and LCP incident wavefronts, i.e., $S_R =$ $(I_{\bar{R}} - I_R)/(I_{\bar{R}} + I_R)$ and $S_L = (I_{\bar{L}} - I_L)/(I_{\bar{L}} + I_L)$. Specifically, the readout signal $S_R$ computed from the photocurrents $I_R$ and $I_{\bar{R}}$ of each superpixel is zero, positive, or negative if the local angle of incidence θ of the RCP component is zero, positive, or negative, respectively, regardless of the total incident intensity and of the direction of propagation of the LCP component (and similarly for $S_L$ with the two states of circular polarization interchanged). Since local angle of incidence is proportional to transverse phase gradient, the resulting edge-enhanced images can then be numerically inverted to reconstruct the underlying RCP and LCP phase distributions.

To substantiate the phase imaging capabilities of our experimental samples in the configuration of Fig. 5, we use a computational imaging model based on the measured angular response maps, following our prior work with plasmonic diffractive photodetectors [4, 6]. This model is based on a Fourier decomposition of the light incident on the sensor array into plane waves propagating along different directions, which are detected by each pixel according to its angular response [52]. In the PIMSs of the present work, the RCP and LCP components of each plane wave are captured (i.e., converted into SPPs which eventually contribute to the photocurrent) with different transfer functions $t_{RCP}(\mathbf{k})$ and $t_{LCP}(\mathbf{k})$ because of the chiral nature of these devices. The magnitudes of these polarization-dependent transfer functions can be obtained from the experimental results of Fig. 3, where the input light is a plane wave of either circular polarization X = RCP or LCP, and the measured responsivity map $\mathcal{R}_X(\mathbf{k})$ is therefore proportional to $|t_X(\mathbf{k})|^2$. In a sensor array illuminated with arbitrary polarization and field distribution, the photocurrent of each pixel can then be evaluated as

$$I \propto \left| \mathcal{F}^{-1}\left\{ \sqrt{\mathcal{R}_{RCP}(\mathbf{k})} e^{i\Delta(\mathbf{k})} E_{RCP}(\mathbf{k}) + \sqrt{\mathcal{R}_{LCP}(\mathbf{k})} E_{LCP}(\mathbf{k}) \right\} \right|^2. \qquad (3)$$

Here $\mathcal{F}^{-1}$ denotes the inverse spatial Fourier transform evaluated at the pixel location, $\Delta(\mathbf{k})$ is the phase difference between the two transfer functions $t_{RCP}(\mathbf{k})$ and $t_{LCP}(\mathbf{k})$, and $E_{RCP}(\mathbf{k})$ and $E_{LCP}(\mathbf{k})$ are the Fourier transforms of the incident RCP and LCP optical fields on the sensor array. Physically, the two terms in the curly brackets of eq. (3) correspond to the SPPs excited by the Fourier components of in-plane wavevector **k** of the RCP and LCP incident waves, which add up coherently to one another before being collected at the slits. To estimate $\Delta(\mathbf{k})$, we have measured the angle-resolved photocurrent of our experimental samples under s and p linearly polarized illumination, and then we have used eq. (3) [with the data of Fig. 3 for $\mathcal{R}_{RCP}(\mathbf{k})$ and $\mathcal{R}_{LCP}(\mathbf{k})$] to fit the resulting responsivity curves. This procedure (described in more detail in Supplementary Material, Section S4) yields accurate fits with $\Delta \approx 43°$ across the pupil-function passband of the envisioned microscope.

Representative computational imaging results based on this model are shown in Fig. 6, where we consider the PIMS array of Fig. 5 combined with a telecentric imaging system with 20× magnification and 0.8 numerical aperture. The incident light used in the calculations consists of a superposition of RCP and LCP components with equal magnitudes and different phase profiles $\varphi_{RCP}(\mathbf{r})$ and $\varphi_{LCP}(\mathbf{r})$ shown on the left-hand side of Fig. 5. Specifically, both $\varphi_{RCP}$ and $\varphi_{LCP}$ oscillate as a function of x with the same period (14 μm) but different amplitudes (0.3 and 0.2 rad, respectively) and opposite polarity. These field distributions allow for a simple illustration of our devices' phase imaging capabilities. At the same time, similar patterns are measured in samples consisting of alternating domains of equal and opposite circular birefringence, such as banded spherulites of various molecular crystals including aspirin [15].

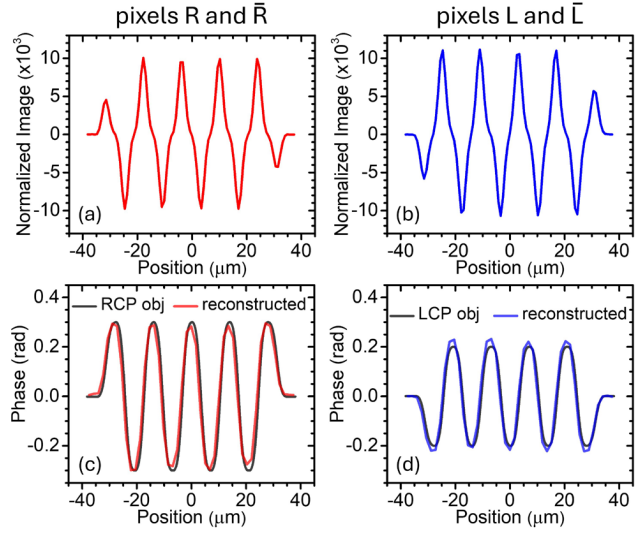

**Figure 6.** Chiral phase imaging simulation results. (a) Horizontal line cut through the middle of the differential-phase-contrast image $S_R(\mathbf{r})$ recorded by devices R and $\bar{\text{R}}$ for the phase object of Fig. 5. (b) Horizontal line cut of the differential-phase-contrast image $S_L(\mathbf{r})$ simultaneously recorded from the readout signals of devices L and $\bar{\text{L}}$. (c) Phase profile $\varphi_{RCP}(\mathbf{r})$ of the RCP component of the incident light (grey trace) and computationally reconstructed image (red trace). (d) Phase profile $\varphi_{LCP}(\mathbf{r})$ of the LCP component of the incident light (grey trace) and computationally reconstructed image (blue trace).

Figures 6(a) and 6(b) display horizontal line cuts of the differential-phase-contrast images $S_R(\mathbf{r})$ and $S_L(\mathbf{r})$ computed with the model of eq. (3). The line cuts of the corresponding phase profiles $\varphi_{RCP}(\mathbf{r})$ and $\varphi_{LCP}(\mathbf{r})$ are shown by the grey lines in Figs. 6(c) and 6(d). Consistent with expectations, at any location where $\varphi_{RCP}(x)$ increases (decreases) with x [i.e., where the incoming RCP component is deflected in the positive (negative) x direction], a peak (dip) is introduced in $S_R(x)$ in Fig. 6(a). Similar considerations apply to $\varphi_{LCP}(x)$ and $S_L(x)$ in Fig. 6(b). Finally, to reconstruct the two phase objects $\varphi_{RCP}(\mathbf{r})$ and $\varphi_{LCP}(\mathbf{r})$ from $S_R(\mathbf{r})$ and $S_L(\mathbf{r})$ we have adapted a numerical inversion algorithm that is commonly used with phase imaging systems based on sequential oblique illumination [53] (see Supplementary Material, Section S5). The horizontal line cuts of the reconstructed phase profiles are plotted in Figs. 6(c) and 6(d), showing good agreement with the original patterns (the full reconstructed images are also displayed on the right-hand side of Fig. 5). These results indicate that the measured angular response characteristics of our devices provide the required sensitivity to phase contrast and polarization selectivity for circular-polarization-resolved wavefront sensing. Further improvements could be obtained with PIMSs featuring sharper responsivity peaks, e.g., by minimizing the surface roughness in the metasurface Au layer or by using additional NP geometries that can provide more uniform scattering efficiency across the array.

## 3. Discussion

We have developed a new class of multifunctional photodetectors that can selectively measure the phase gradient of either circular polarization component of any incident wave. The design and operation of these devices involves multiple ideas from plasmonics and nanophotonics, including chiral wavefront shaping with gap-plasmon and PB-phase metasurfaces, extraordinary optical transmission, and spin-momentum locking in SPPs. An order-of-magnitude polarization selectivity with respect to phase gradient is demonstrated by photocurrent measurements with our experimental samples. Computational imaging simulations based on the measured device characteristics illustrate the ability of these chiral PIMSs to image circular birefringence with a particularly simple single-shot measurement protocol in a highly compact setup. In turn, this capability is of interest to add miniaturization and speed in applications ranging from pharmaceutical drug development to fundamental studies of live cells and biomedical diagnostics.

Finally, we highlight additional functionalities that may be enabled by the same devices in suitably designed sensor arrays. First, the four photocurrent signals recorded by each superpixel in the configuration of Fig. 5 may be combined and analyzed to reconstruct not only the phase gradients but also the magnitudes of the RCP and LCP components, for the simultaneous imaging of circular birefringence and circular dichroism. Second, larger superpixels may be constructed including additional replicas of devices R and L rotated by ±90°. The resulting sensor array would allow for the measurement of wavefront distortions along all transverse directions with comparable sensitivity (albeit at the expense of decreased spatial resolution). In contrast, in the configuration of Fig. 5 deflections along the y directions produce smaller changes in the photocurrent signals due to the vertical orientation of the responsivity peaks. Finally, the same devices could be combined with the diffractive PIMSs reported in our prior work [6], which only detect linearly polarized light. The photocurrent signals of the resulting superpixels could then be decoded numerically to retrieve the spatial distribution of the incident intensity, state of polarization, and phase gradient, in a single measurement with standard imaging optics. The end result would be a uniquely advanced imaging system that could dramatically increase our ability to visualize objects in low-contrast environments and to extract maximum information from the resulting images. Such a system could therefore find and enable new applications well beyond chiral biochemical sensing, extending into areas such as autonomous navigation, computer vision, and optical information processing.

## 4. Materials and Methods

**Design simulations.** All the simulations presented in this work were carried out with the Ansys-Lumerical FDTD Solutions software package. The two angular response traces of Fig. 1(b) were generated by computing the transmission through the entire metasurface of, respectively, an RCP and LCP diffractive plane wave incident from the air above as a function of angle of incidence $\theta$. The reflection amplitude and phase values displayed in Fig. 2(b) were computed via simulations of a single NP oriented along the axes of the metasurface array [i.e., with $\alpha = 0$ in Fig. 2(a)], using periodic boundary conditions and illumination at normal incidence with linear polarization along the NP axes.

**Device fabrication.** The experimental samples were fabricated on undoped (100) Ge substrates. The fabrication process includes electron-beam evaporation and plasma-enhanced chemical vapor deposition for the Au and $SiO_2$ films, respectively, electron-beam lithography for the NP array, and focused ion beam milling for the slits. Each completed device was mounted on a copper block and wire-bonded to two Au-coated ceramic plates.

**Device characterization.** The measurement results presented in Figs. 3 and 4 were collected with a custom-built optical goniometer setup that allows varying both polar $\theta$ and azimuthal $\phi$ illumination angles. The device under study is biased with a 1-V dc voltage and illuminated with 0.5-mW light from a diode laser (modulated at 1 kHz, so that the photocurrent can be separated from the dark current using a bias tee and lock-in amplifier). The laser light is delivered to the sample with a polarization-maintaining fiber mounted in a cage system, which contains a polarizer, a half-wave plate, and a quarter-wave plate used to generate the desired states of polarization.

## Author Statements

## Acknowledgements

The FDTD simulations were performed using the Shared Computing Cluster facility at Boston University. Some of the fabrication tasks were carried out at the Center for Nanoscale Systems of Harvard University.

**Corresponding author**

Email: rpaiella@bu.edu

**Research funding**

This work was supported by the National Science Foundation under Grant # ECCS 2139451.

**Author contributions**

All authors have accepted responsibility for the entire content of this manuscript and approved its submission.

**Conflict of interest**

Authors state no conflicts of interest.

**Data availability**

The datasets generated and/or analyzed during the current study are available from the corresponding author upon reasonable request.